\documentclass[reprint,aps,amsmath,amssymb,amsfonts]{revtex4-1}
\usepackage{graphicx}
\usepackage{dcolumn}
\usepackage{bm}
\usepackage{txfonts}
\usepackage{multirow}

\begin{document}

\title{Spectroscopic signature of trimer Mott insulator and charge disproportionation in BaIrO$_3$}

\author{R.~Okazaki$^{1,\ast}$}
\author{S.~Ito$^{2}$}
\author{K.~Tanabe$^{2,\dagger}$}
\author{H.~Taniguchi$^{2}$}
\author{Y.~Ikemoto$^{3}$}
\author{T.~Moriwaki$^{3}$}
\author{I.~Terasaki$^{2}$}

\affiliation{$^1$Department of Physics, Faculty of Science and Technology, Tokyo University of Science, Noda 278-8510, Japan}
\affiliation{$^2$Department of Physics, Nagoya University, Nagoya 464-8602, Japan}
\affiliation{$^3$SPring-8, Japan Synchrotron Radiation Research Institute (JASRI), Sayo, Hyogo 679-5198, Japan}

\begin{abstract}
We have measured the reflectivity spectra of the barium iridate $9R$ BaIrO$_3$, 
the crystal structure of which 
consists of characteristic Ir$_3$O$_{12}$ trimers.
In the high-temperature phase above the transition temperature $T_c\simeq180$~K, 
we find that
the optical conductivity involves two temperature-dependent optical transitions
with an ill-defined Drude response.
These features are reminiscent of the optical spectra in the organic dimer Mott insulators,
implying
a possible emergence of an unusual electronic state named trimer Mott insulator in BaIrO$_3$,
where the carrier is localized on the trimer owing to the strong Coulomb repulsion.
Along with a pronounced splitting of the phonon peak observed below $T_c$, 
which is a hallmark of charge disproportionation,
we discuss a possible phase transition from the trimer Mott insulator to a 
charge-ordered insulating phase in BaIrO$_3$.
\end{abstract}

\maketitle

\section{Introduction}

Recent observations of exotic electronic states in $4d$- and $5d$-electron systems,
such as unusual orbital ordering in ruthenates \cite{Kimber2009,Kunkemoller2015,Jain2017} and 
the spin-orbit Mott insulator in iridates \cite{Kim2008,Kim2009,Jackeli2009,Kitagawa2018},
have emphasized the emergent physics of correlated electron systems.
In contrast to well-localized $3d$ orbital, 
$4d$ and $5d$ electrons 
have extended electronic wave functions,
making the Coulomb repulsion comparable to the bandwidth.
These energy scales also compete with 
spin-orbit interactions, 
leading to intriguing spin-orbital-entangled electronic states in those materials.

Another important aspect of the $4d$- and $5d$-electron systems with 
spatial extent of the $d$ orbitals is 
close coupling with the lattice degrees of freedom.
A representative unit structure in the transition-metal oxides is 
MO$_6$ octahedron (M being transition-metal ion), 
which tends to form a corner-shared network like the perovskite structure 
as seen in various $3d$ oxides.
On the other hand,
in $4d$ and $5d$ oxides,
the network structures made of edge- and face-sharing octahedra
also become stable
owing to the direct hybridization among the extended $d$ orbitals.
Interestingly, such compounds exhibit anomalous phase transitions 
related to a molecular orbital formation of $d$ orbitals,
such as a metal-insulator transition in Li$_2$RuO$_3$ driven by
a dimerization of edge-sharing RuO$_6$ octahedra \cite{Miura2007,Terasaki2015}
and an antiferromagnetic transition in Ba$_4$Ru$_3$O$_{10}$ related to
a molecular orbital formation in Ru$_3$O$_{12}$ trimers 
made of face-shared octahedra \cite{Igarashi2013,Igarashi2015}.

$9R$ phase BaIrO$_3$ is particularly unique
among them, since it exhibits an intriguing quasi-one-dimensional crystallographic form and 
an unconventional phase transition at $T_c \simeq 180$~K,
below which
a weak ferromagnetism and a gap opening simultaneously set in \cite{Siegrist1991,Powell1993,Lindsay1993,Cao2000,Brooks2005}.
The crystal structure is composed of Ir$_3$O$_{12}$ trimers,
made up of three face-sharing IrO$_{6}$ octahedra as depicted in the inset of Fig.~1(a) \cite{Siegrist1991}.
Note that
the Ir-Ir distance in the trimer is shorter than that of Ir metal,
indicating a strong hybridization of the intra-trimer $5d$ orbitals.
The corner-sharing trimers form a zig-zag chain along the $c$ axis \cite{Siegrist1991}.
The quasi-one-dimensional nature also appears in the anisotropic transport properties \cite{Cao2000,Nakano2006}.

The issue to be addressed is how the structural feature correlates with the origin of the phase transition at $T_c$.
Various measurements including 
resistivity \cite{Cao2000},
magentoresistance \cite{Cao2004},
thermopower \cite{Kini2005,Nakano2006},
optical spectroscopy \cite{Cao2000}, and
photoemission \cite{Maiti2005},
suggest a (partial) gap opening in the density of states below $T_c$,
indicating a charge-density-wave (CDW) formation
as is expected from the low-dimensional crystal structure.
This is supported by tight-binding and local-spin-density approximation
calculations \cite{Whangbo2001, Maiti2006},
which also suggest an itinerant ferromagnetism within a Stoner picture.
On the other hand, the CDW model is not straightforward 
because
the nonmetallic transport behaviors are reported both above and below $T_c$ \cite{Cao2000,Kini2005,Nakano2006},
and to our knowledge, 
there is no experimental evidence for the appearance of the superlattice reflections 
accompanying the CDW  formation below $T_c$ \cite{Mori_JPS,Terasaki2016}.
Furthermore, the significance of spin-orbit interaction is also suggested \cite{Marco2010,Ju2013},
complicating the underlying mechanism of the phase transition.

In this study, we present the detailed temperature variation of the optical conductivity spectra of BaIrO$_3$
to grasp the electronic structure in a wide energy range.
Above $T_c$, we find that
the optical conductivity spectra
possess a qualitative similarity to that of the organic dimer Mott insulators,
in which the correlated carrier is localized on the dimerized molecules to act as a Mott insulator.
Thus this result indicates a molecular orbital formation of $5d$ orbitals in the Ir$_3$O$_{12}$ trimers
and 
a possible emergence of an unusual electronic state named trimer Mott insulator,
where the carrier is localized on the trimer by the Coulomb repulsion, 
as is suggested by recent structural investigation \cite{Terasaki2016}.
Below $T_c$, the optical phonon peaks further split,
indicating an occurrence of a charge disproportionation.
Based on the present results, 
we discuss an inter-trimer charge degree of freedom 
to drive a phase transition from the trimer Mott to a charge order insulator.

\begin{figure}[t]
\begin{center}
\includegraphics[width=1\linewidth]{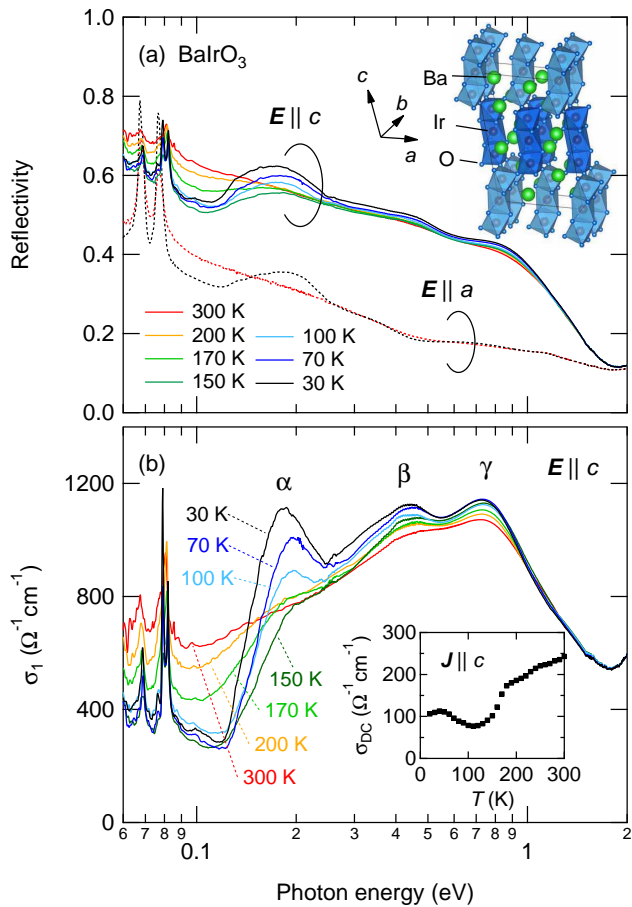}
\caption{(Color online).
(a) Reflectivity spectra measured at several temperatures with polarization parallel to
the $c$ axis (solid curves) and the $a$ axis (dotted curves).
The inset depicts the crystal structure of $9R$ BaIrO$_3$.
Light- and dark-blue colors show the inequivalent Ir$_3$O$_{12}$ trimers.
(b) Optical conductivity spectra measured with polarization parallel to
the $c$ axis.
The inset shows the temperature dependence of the dc conductivity along the $c$ axis  taken from
Ref. \onlinecite{Nakano2006}.
}
\end{center}
\end{figure}

\section{Experiments}

Single-crystalline samples of BaIrO$_3$ 
with a typical dimension of $0.3\times0.3\times1$~mm$^3$ 
were grown by the self-flux method \cite{Nakano2006}.
Powders of IrO$_2$ (99.9\%) and BaCl$_2$ (99.9\%)
were mixed in a molar ratio of $1:10$.
The mixture was put in an alumina crucible and 
heated up to 1523~K in air with a heating rate of 200~K/h.
After keeping 1523~K  for 12~h, 
it was slowly cooled down with a rate of 1~K/h, 
and at 1123~K the power of the furnace was switched off.
The crystal axis was determined by Laue method.

The polarized reflectivity spectra were measured using a 
Fourier transform infrared spectrometer (FTIR) 
with a synchrotron radiation (SR) light at BL43IR, SPring-8, Japan for energies between 60 meV and 1.0 eV \cite{Kimura2013},
an FTIR (FTIR-6000 and IRT-5000, JASCO Corp.) for energies between 0.2 eV and 1.4 eV, and
a grating spectrometer for energies between 1.2 eV and 2.5 eV.
For the infrared range from 60 meV to 1.4 eV, 
we used a helium flow-type refrigerator and
the sample was fixed with a silver paste on the cold head. 
Here we used a standard gold overcoating technique for measuring the reference spectrum at each temperature.
For the visible range, we have only performed room-temperature measurement and 
used a SiC for the reference \cite{Pressley_book}.
The complex optical conductivity is obtained from the Kramers-Kronig (KK) analysis.
We used a constant extrapolation of the reflectivity up to 6 eV and subsequently employed
a standard extrapolation of $\omega^{-4}$ dependence.
In low energies, we extrapolated the reflectivity using several methods,
but the influence on the peak shapes and positions is negligible for discussion.

\begin{figure}[t]
\begin{center}
\includegraphics[width=7.5cm]{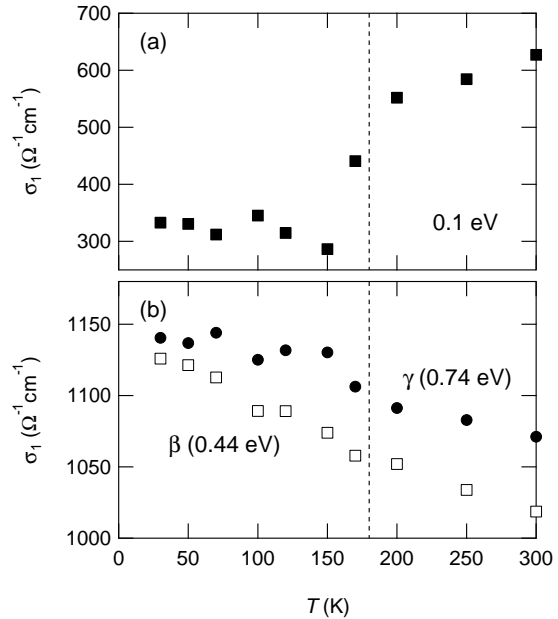}
\caption{
(a) Temperature dependence of the low-energy optical conductivity at 
$\hbar\omega=0.1$~eV.
The dashed line represents the transition temperature $T_c\simeq180$~K.
(a) Temperature variations of the optical conductivity at 
the $\beta$ peak (open squares) and the $\gamma$ peak (filled circles).
}
\end{center}
\end{figure}

\section{Results and discussion}

\subsection{Reflectivity and optical conductivity}

Figure 1(a) shows the temperature variation of the reflectivity spectra.
The reflectivity measured with the polarization 
parallel to the $c$ axis (solid curves) is significantly larger than that for $a$-axis (dotted curves), 
consistent with the transport anisotropy \cite{Cao2000,Nakano2006}.
Note that earlier optical experiments used unpolarized light above 0.74 eV \cite{Cao2000},
which may result in worse KK transformation.
Hereafter we focus on the optical data for the conducting $c$ axis.

The reflectivity spectra exhibit a considerable temperature dependence.
Above $T_c$, the optical phonon structures below 0.1 eV are unscreened and 
the low-energy reflectivity decreases with lowering temperature, indicating a nonmetallic high-temperature phase.
In the infrared range, we notice three temperature-dependent peak structures 
around 0.2, 0.4, and 0.7 eV, which will be discussed in the optical conductivity spectra.
On the other hand, the temperature dependence becomes small and converges around 1 eV.
Thus we used the room-temperature data for the reflectivity spectra of all temperatures above 1.4 eV as mentioned
in the previous session
and the discussion on the temperature variation will be restricted only below 1 eV. 

Figure 1(b) depicts the real part of the optical conductivity spectra $\sigma_1$.
We find three temperature-dependent peak structures
labelled as $\alpha$ ($\sim 0.2$ eV), $\beta$ ($\sim 0.4$ eV), $\gamma$ ($\sim 0.7$ eV)
in the infrared range, and 
the temperature variations of the optical conductivity at several characteristic photon energies
are shown in Figs. 2.
In contrast to the present result,
the optical conductivity of $J_{\rm eff}=1/2$ Mott insulator Sr$_2$IrO$_4$ 
involves two characteristic peaks below 1 eV \cite{Moon2009},
implying intrinsically different electronic structures among these iridates.

\subsection{High-temperature phase above $T_c$}

First we discuss the high-temperature electronic state.
As shown in Figs. 1(b) and 2(a), 
above $T_c$,
the optical conductivity in the low-energy region around 0.1 eV is 
gradually suppressed with lowering temperature.
Moreover, as shown in the inset of Fig. 1(b),
the DC conductivity $\sigma_{\rm DC}$ \cite{Nakano2006},
which equals to $\sigma_1(\omega\to0)$,
is smaller than 
the optical conductivity in the low-energy range. 
These results clearly demonstrate a nonmetallic nature in the high-temperature phase of BaIrO$_3$,
consistent with the nonmetallic transport \cite{Cao2000,Kini2005,Nakano2006} 
and the localized electronic state \cite{Maiti2005}
even above $T_c$.
The temperature dependence of the optical conductivity at 0.1 eV [Fig. 2(a)] is qualitatively similar to 
that of DC conductivity [inset of Fig. 1(b)], 
indicating that 
the transport properties are strongly affected by 
the electronic structure in a wide energy range.

To further understand the insulating nature above $T_c$, we then discuss the origins of the
temperature-dependent peaks observed in the infrared range.
Now we consider
a localized picture of Ir $5d$ electrons proposed in Ref. \onlinecite{Terasaki2016}.
As schematically drawn in Figs. 3(a) and (b), 
one Ir$^{4+}$ ion has the electronic configuration of $5d^5$ 
in the triply-degenerate $t_{2g}$ orbital, which
further splits into higher $a_{1g}$ and lower doubly-degenerate $e'_{g}$ orbitals owing to a trigonal distortion.

\begin{figure}[t]
\begin{center}
\includegraphics[width=1\linewidth]{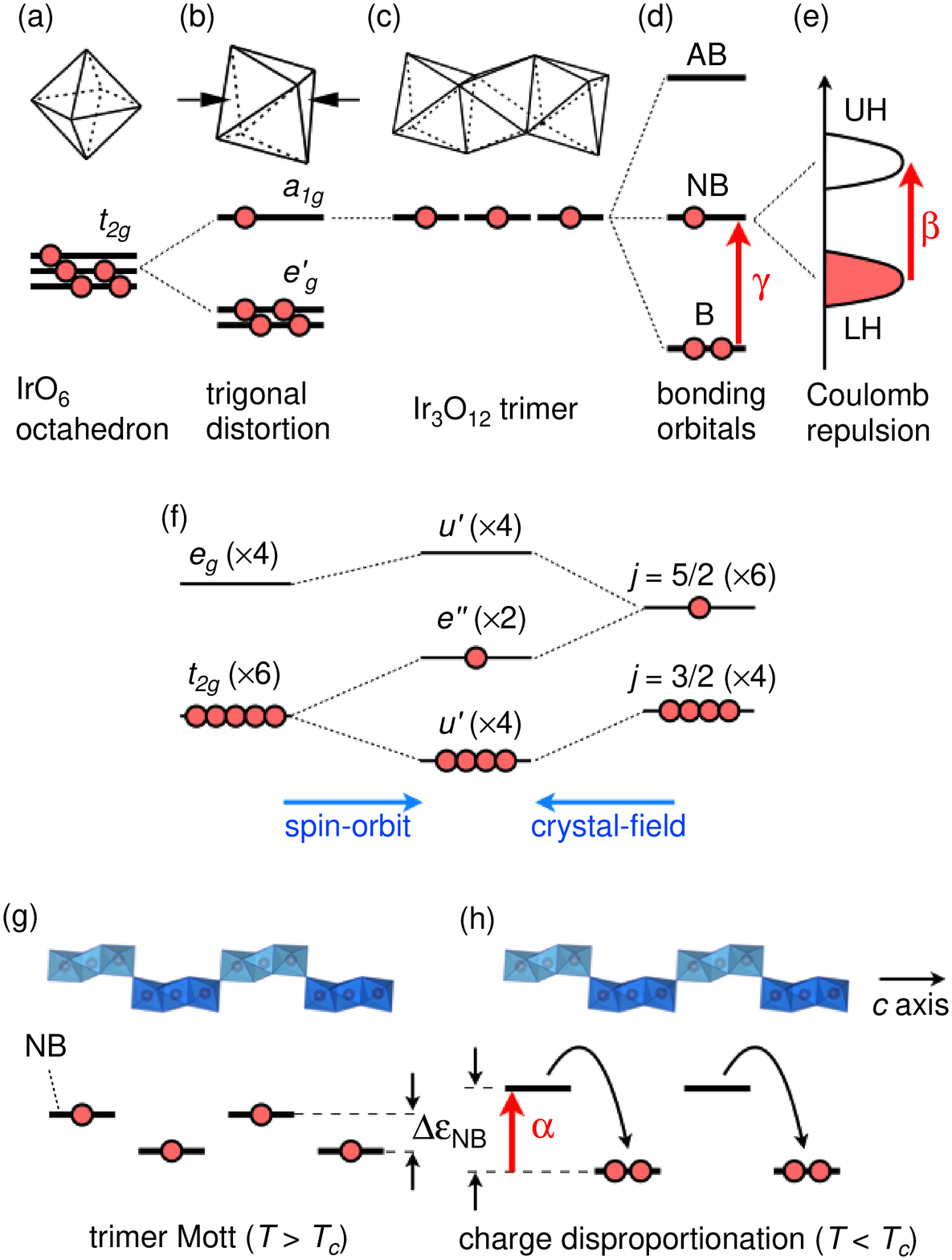}
\caption{(Color online).
Schematic energy diagrams for the electronic state of BaIrO$_3$ constructed from a localized picture \cite{Terasaki2016}.
The red arrows labelled as $\alpha$, $\beta$, $\gamma$ correspond to the optical transitions shown in Fig. 1(b). 
Energy diagrams for Ir $5d$ electrons in (a) IrO$_6$ octahedron, (b) trigonally distorted octahedron, (c) Ir$_3$O$_{12}$ trimer.
(d) In the trimer, three $a_{1g}$ orbitals form the bonding (B), non-bonding (NB), and anti-bonding (AB) orbitals.
(e) Under strong correlation effect, the half-filled NB orbitals split into lower-Hubbard (LH) and upper-Hubbard (UH) bands.
(f) Energy levels for Ir$^{4+}$ under spin-orbit and crystal-field interactions \cite{Marco2010}.
(g) Trimer Mott insulating state realized above $T_c \simeq 180$~K. 
The energy difference of NB orbitals $\Delta\varepsilon_{\rm NB}$ between two inequivalent trimers is relatively small.
(h) Charge disproportionation below $T_c$.
}
\end{center}
\end{figure}

An important point is that BaIrO$_3$ consists of the Ir$_3$O$_{12}$ trimers in which the Ir-Ir distance is fairly short.
We then consider a molecular orbital formation between three $a_{1g}$ orbitals within a trimer,
and 
obtain bonding (B), non-bonding (NB), and anti-bonding (AB) orbitals as shown in Figs. 3(c) and (d).
In this case, we find one electron in the NB orbital, and
this state can be regarded as half-filling if we identify one trimer as one structural unit.
In this picture, we may assign the highest $\gamma$ peak as the transition between
B and NB orbitals (trimer peak), 
which is good agreement with the estimation of $\sim 1$~eV by the band calculation \cite{Ju2013}.
Also, the transition between NB and AB orbitals, in which the energy difference is close to that of the B-NB transition, 
 is also included in the $\gamma$ peak.
Note that the transition from B to AB orbitals is prohibited in a simple dipole transition since both orbitals have even parity.
Furthermore, as depicted in Fig. 3(e),
in the presence of the Coulomb interaction, this half-filled state can become a Mott insulator, 
which can explain the insulating behaviors mentioned above, 
and 
the transition between the lower- and upper-Hubbard bands
may be assigned as the $\beta$ peak.

Strong spin-orbit interaction, 
essential for understanding electronic states of iridates \cite{Kim2008,Kim2009},
can be taken in the present model.
As depicted in Fig. 3(f), under spin-orbit and crystal-field interactions, 
the $5d$ states split into highest quadruply-degenerate $u'$, 
doubly-degenerate $e''$, and lowest quadruply-degenerate $u'$ levels, and
the highest-occupied level is the $e''$ level as is suggested by x-ray absorption study \cite{Marco2010}.
However, even in such a case, the $e''$ level instead of the $a_{1g}$ level may form molecular orbitals in the trimer,
and thus the situation to realize the half-filled NB state is unaltered.

The present insulating state can be called trimer Mott insulator \cite{Terasaki2016},
in analogy to the dimer Mott insulator, in which 
a correlated carrier is localized on the dimerized organic molecules to act as a Mott insulator \cite{Seo2006}.
Indeed, the optical conductivity of BaIrO$_3$ has a similarity to that of the dimer Mott insulator:
It also involves a transition between bonding and anti-bonding orbitals in a dimer (dimer peak) and
a transition between Hubbard bands \cite{Faltermeier2007}.
Moreover, the intensities of these optical transitions increase with lowering temperature \cite{Sasaki2004},
also similar to the enhancements of $\beta$- and $\gamma$-peak intensities with cooling
as shown in Fig. 2(b).
Note that these intensities contrastingly decrease with lowering temperature to enhance the Drude weight 
in correlated metallic state \cite{Sasaki2004}.
Thus the present results of the optical conductivity above $T_c$ may capture the essential features of 
a trimer Mott insulator state in the high-temperature phase of BaIrO$_3$.

A schematic picture of the trimer Mott insulator state above $T_c$ is drawn in Fig. 3(g).
The correlated carrier is localized at each NB orbital, which is formed by the $a_{1g}$ or the $e''$ levels.
Note that there are two crystallographically-inequivalent trimers \cite{Siegrist1991},
and the energy difference between the NB orbitals of these trimers $\Delta\varepsilon_{\rm NB}$ may be small above $T_c$.

\subsection{Low-temperature phase below $T_c$}

Having the above arguments in mind,
let us discuss the low-temperature electronic state below $T_c$.
The distinct feature below $T_c$ is 
an enhancement of the $\alpha$ peak shown in Fig. 1(b), 
which is reminiscent of
those in CDW compounds \cite{Degiorgi1994} and 
in charge order systems \cite{Ivek2011,Okazaki2013},
indicating that a similar electronic state emerges in the low-temperature phase of BaIrO$_3$.
Here the trimer Mott insulator state discussed above is unstable against 
the charge order (charge disproportionation), 
because one must have
finite energy difference between the two NB levels of crystallographically-inequivalent trimers ($\Delta\varepsilon_{\rm NB}\neq0$).
Thus, as shown in Fig. 3(h), if $\Delta\varepsilon_{\rm NB}$ is relatively large, 
a carrier is transferred from the high- to the low-energy NB level, 
leading to a charge disproportionation.
Indeed, recent x-ray diffraction study using synchrotron radiation has revealed that
the two trimers become more inequivalent with lowering temperature \cite{Terasaki2016},
supporting the charge disproportionation scenario.
In this case, the $\alpha$ peak corresponds to the transition between two NB levels with the energy difference $\Delta\varepsilon_{\rm NB}$.
Note that the low-temperature enhancement of the $\alpha$ peak is also observed clearly in the reflectivity measured with the polarization 
parallel to the $a$ axis [Fig. 1(a)],
consistent with the gap opening for the charge disproportionation below $T_c$.

\begin{figure}[t]
\begin{center}
\includegraphics[width=1\linewidth]{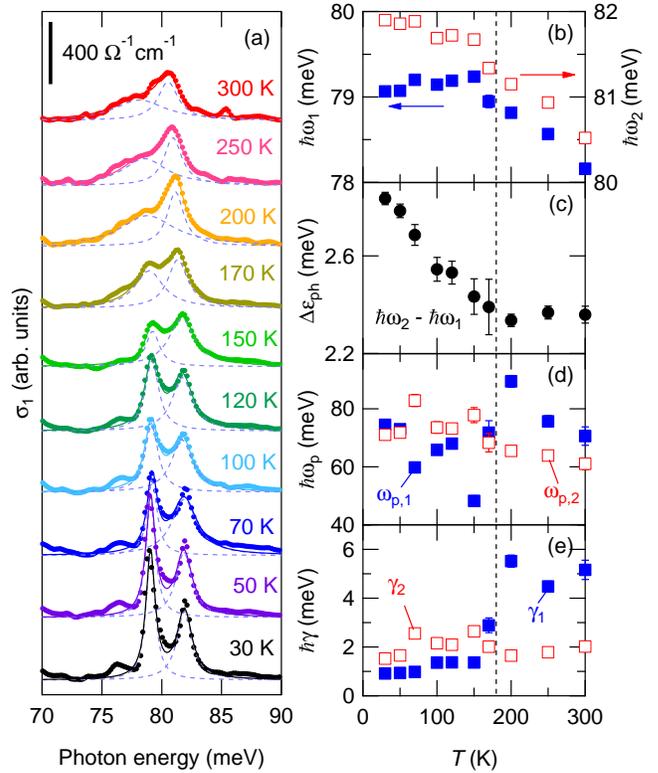}
\caption{(Color online).
(a) Optical phonon spectra measured at several temperatures
with polarization parallel to the $c$ axis.
The experimental data are fitted using two Lorentz oscillators
shown as the dashed curves for each temperature.
The sum of two oscillators is represented as the solid curves.
(b-e) Temperature variations of the fitting parameters;
(b) the resonance energies at around $\hbar\omega_1\simeq79$~meV (solid squares, left axis)
and $\hbar\omega_2\simeq81$~meV (open squares, right axis),
(c) the difference of the resonance energies $\Delta\varepsilon_{\rm ph} = \hbar\omega_2 - \hbar\omega_1$,
(d) the plasma frequencies, and 
(e) the damping factors.
The vertical dashed line represents the transition temperature $T_c\simeq180$~K.
}
\end{center}
\end{figure}

In order to examine this scenario, 
we focus on the phonon spectra, which are sensitive to the charge disproportionation \cite{Takahashi2006}.
As plotted in Fig. 4(a),
we find two temperature-dependent peaks around 80 meV,
which may be assigned as stretching modes of IrO$_6$ \cite{Moon2009},
although the thorough assignment is incomplete.
We then try to fit these spectra with two Lorentz oscillators,
\begin{equation}
\sigma_1(\omega) = \varepsilon_0\omega\sum_{k=1}^{2}\omega^2_{p,k}
\frac{\gamma_k\omega}{(\omega_k^2-\omega^2)^2+\gamma_k^2\omega^2},
\end{equation}
where $\varepsilon_0$ is the electric constant and 
$\omega_{p,k}$, $\omega_{k}$, and $\gamma_{k}$ are 
the plasma frequency, the resonance frequency, and the damping factor of 
the $k$th oscillator, respectively.
In Fig. 4(a), 
the fitting results of each oscillator and the sum are shown by the dashed and the solid curves, respectively.
The two-oscillators fitting well reproduces the experimental results below and above $T_c$,
indicating no translational symmetry breaking at $T_c$, 
consistent with the earlier diffraction studies \cite{Mori_JPS,Terasaki2016}.
Note that a very small peak around 76 meV, which is not taken account of in this fitting, 
does not affect the results.

Figures 4(b) and 4(c) show the temperature variations of 
the resonance energies $\hbar\omega_k$ $(k=1,2)$ and 
the difference $\Delta\varepsilon_{\rm ph} = \hbar\omega_2 - \hbar\omega_1$, respectively.
Below $T_c$, $\Delta\varepsilon_{\rm ph}$ dramatically increases with decreasing temperature, 
similar to the phonon peak splitting in the charge order phase 
in oxides \cite{Katsufuji1996}
and organic systems \cite{Tanaka2008,Okazaki2013}.
Whereas a critical difference is that 
the charge-order compounds exhibit translational symmetry breaking 
while the present iridate does not,
the increase of $\Delta\varepsilon_{\rm ph}$ below $T_c$ indeed suggests 
that a charge imbalance is pronouncedly enhanced in the low-temperature phase.
Here there are four crystallographically-inequivalent Ir sites in the unit cell 
(i.e. central and edge Ir sites for two inequivalent trimers), 
but a NB nature with no central weight of the wave function indicates that 
the edge IrO$_6$ octahedra are more sensitive to the amount of charge rather than the central one.
Indeed, the theoretical calculation has shown that 
the density of states of the central Ir ions in the trimers
is suppressed at the Fermi energy \cite{Whangbo2001},
which is naturally understood within the NB nature.
Therefore the enhancement of $\Delta\varepsilon_{\rm ph}$ may reflect
a difference of charge amounts 
between the edge Ir sites of the inequivalent trimers, 
indicating 
an inter-trimer charge disproportionation shown in Fig. 3(h).
We note that, in opposite to this charge disproportionation among the NB orbitals,
the $\beta$ peak, which reflects the feature of the trimer Mott state in the NB orbital, is negligibly affected and enhanced even at low temperatures as shown in Fig. 2(b).
This possibly implies that 
the amount of inter-trimer charge difference is small, similarly to
some charge-ordered organic compounds with small difference in the site charge $\Delta\rho < 0.1e$ \cite{Yakushi2012}. 
Nevertheless, the concurrent observation of the 
phonon peak splitting and the $\alpha$ peak growth below $T_c$ 
has a 
striking similarity to that of charge-ordered organic salts \cite{Ivek2011,Okazaki2013,Tanaka2008}, 
strongly indicating an emergence of the charge disproportionation.

Temperature variations of 
the plasma frequencies $\hbar\omega_{p,k}$ and 
the damping factors $\hbar\gamma_k$ $(k=1,2)$
are shown in Fig. 4(d) and 4(e), respectively.
While these data are scattering, we find that 
$\hbar\gamma_{1}$ is anomalously enhanced above $T_c$,
as is clearly seen in the high-temperature phonon spectra in Fig. 4(a).
Such a broadening is also reported in the high-temperature phase of the 
organic charge order systems and is considered as a fluctuation of the charge order \cite{Yue2010}.
Thus the phonon peak broadening at high temperatures in BaIrO$_3$ possibly indicates 
a fluctuation or a precursor of the charge disproportionation.

We finally mention the low-energy optical conductivity,
which is suppressed but 
a finite weight remains even below $T_c$ [Figs. 1(b) and 2(a)].
This result suggests a pseudogap formation,
as is consistent with the optical and photoemission studies \cite{Cao2000,Maiti2005}.
Such a pseudogap behavior has also been observed in the optical conductivity of BaRuO$_3$,
which possesses a Ru$_3$O$_{12}$ trimer \cite{Lee2001}.
We infer that, since the organic dimer Mott insulator competes 
with the charge order phase \cite{Okazaki2013},
the trimer Mott insulator and charge-transferred state proposed above may compete with 
each other to generate such a pseudogap behavior.

\section{Summary}

In summary, the optical conductivity measurements on BaIrO$_3$ single crystals 
shed light on the unconventional phase transition at $T_c\simeq180$~K in this trimer-based compound.
Above $T_c$, the optical conductivity spectra 
clearly show an insulating behavior and 
a qualitative similarity to that of the organic dimer Mott insulators,
indicating an importance of the molecular orbital formation among $5d$ orbitals in the trimer
and 
a possible emergence of the trimer Mott insulator.
Below $T_c$, we find further splitting of the optical phonon peaks,
indicating an occurrence of a charge disproportionation.
The present results emphasize 
an inter-trimer charge degree of freedom  to drive
a phase transition from the trimer Mott to
charge order insulator in BaIrO$_3$, and 
may offer a fascinating playground to investigate the physics of {\it inorganic} molecular solids.

\section*{Acknowledgements}

We thank J. Akimitsu and T. Arima for valuable comments.
The experiments using synchrotron radiation were performed at BL43IR in SPring-8 
with the approvals of JASRI (No. 2014B1116, No. 2015A1189).
This work was supported by the JSPS KAKENHI 
(No. 25610091, No. 26287064, No. 26610099, No. 17H06136, No. 18K13504).

\end{document}